\documentclass[aps,pre,dvipdfm,11pt]{revtex4}
\usepackage{graphicx}


\def\PRL{ Phys. Rev. Lett. }
\def\PRE{ Phys. Rev. E }

\def\RMP{Rev. Mod. Phys. }

\def\BJP{Braz. J. Phys. }

\def\be{\begin{equation}}
\def\ee{\end{equation}}
\def\bea{\begin{eqnarray}}

\def\eea{\end{eqnarray}}
\def\la{\label}
\def\bsea{\begin{subeqnarray}}
\def\esea{\end{subeqnarray}}

\begin{document}
\title{
Double transitions, non-Ising criticality and critical absorbing phase in an interacting monomer-dimer model
on a square lattice}
\author{Keekwon Nam${}^1$, Sangwoong Park${}^2$, Bongsoo Kim${}^1$ and Sung Jong Lee${}^2$}
\affiliation{${}^ 1$ Department of Physics, Changwon National University, 
Changwon 641-773, Korea\\
${}^ 2$ Department of Physics, University of Suwon, Hwaseong 445-743, Korea}
\date{\today}
\begin{abstract}
We present a numerical study on an interacting monomer-dimer model with nearest neighbor 
repulsion on a square lattice, which possesses two symmetric absorbing states.
The model is observed to exhibit two nearby continuous transitions:
the $Z_2$ symmetry-breaking order-disorder transition and the absorbing transition with directed 
percolation criticality.  We find that the symmetry-breaking transition shows a non-Ising critical behavior, 
and that the absorbing phase becomes critical, in the sense that the critical decay of the dimer density 
observed at the absorbing transition persists even within the absorbing phase. 
Our findings call for further studies on the microscopic models and corresponding continuum description 
belonging to the generalized voter university class.
\end{abstract}

\maketitle
\section{Introduction}
Interacting systems often exhibit phase transitions into permanently arrested, absorbing
states in non-equilibrium (NEQ) steady states \cite{marro-dickman,henkel}. 
Intense efforts  have been devoted to classifying these NEQ phase transitions into 
distinct universality classes,  which is one of the most fundamental issues in the NEQ statistical mechanics \cite{odor}.
Although the guiding principles for the classification are still lacking in general, 
symmetry is undoubtedly one of the crucial ingredients in the classification.
In the most prominent universality class known as the directed percolation (DP) class \cite{grassberger1,hinrichsen,takeuchi},  
 systems possess either a single absorbing state or multiple ones not connected by 
any underlying symmetry \cite{janssen,grassberger2}. 
Another newly established class known as the generalized voter (GV) universality class \cite{dornic,hammal}
 encompasses a large variety of models possessing two symmetric absorbing states, 
including statistical models for the social dynamics such as opinion spreading \cite{castellano}.
This universality class embraces most of one dimensional 
models belonging to the so called directed Ising class, such as NEQ kinetic Ising model \cite{menyhard,nam}, 
interacting monomer-dimer (IMD) model \cite{hpark}, and the voter model with spin-exchange \cite{dornic}.

A fundamental and unique feature of the GV class is that, due to the symmetry of the absorbing states,  
the models in the class actually involve two types of phase transitions, namely, 
the $Z_2$ symmetry breaking (SB) order-disorder and the active-absorbing phase transitions.
In one dimension, the coincidence of these two transitions appears 
to be inevitable since the SB should be induced by the absorbing transition. 
 In two dimensions, the models belonging to the GV class are shown to exhibit 
two types of behaviors as follows.
The SB order-disorder transition and the active-absorbing transition
can take place simultaneously at a single critical point \cite{mylee}, where the defect density 
shows a logarithmic decay, as in the original voter model \cite{krapivsky}. 
Close but separate occurrence of these two transitions is also observed \cite{droz}.
Continuum field theoretic descriptions unifying these two types of behavior were formulated
via appropriate Langevin equations for the local order parameter  field \cite{hammal,vazques,castellano2}.

We here present an extensive simulation study on the IMD model with  
nearest neighbor (NN) repulsion on a square lattice. 
We observe that the two transitions are occurring successively at close but clearly distinct  points.  
Moreover, we find that the SB transition reveals a new {\em non-Ising} critical behavior, 
and that the absorbing transition exhibits a DP-critical behavior (due to the $Z_2$-SB), 
that is, the density of active sites, $\rho(t)$, shows a power-law decay in time 
as $\rho(t) \sim t^{-\phi}$ with $\phi \simeq 0.45$, 
in agreement with the corresponding DP exponent in two dimensions. 
Interestingly, we further find that the entire absorbing phase becomes {\em critical} 
in the sense that the same power-law relaxation of $\rho(t)$ persists even deep in the 
absorbing phase.  
In previous works for the microscopic models belonging to the GV class, we feel that the nature of the SB transition was presumed to be 
of pure Ising type, and thus was not carefully studied so far: 
the Binder cumulants associated with the SB transition 
 took the values  $U \simeq 0.56$ \cite{vazques} or  $U \simeq 0.59$ \cite{droz}, 
smaller than the universal Ising value $U_{Ising} \simeq 0.611$ \cite{blote}. 
In contrast, the pure Ising nature of the SB transition in the corresponding continuum Langevin description was reported though explicit 
results are not shown \cite{hammal}.    
Our findings thus call for further studies on these microscopic models belonging to the GV universality class, and on the corresponding 
Langevin equations. Further related discussions are given below. 

\section{ Interacting Monomer-dimer model on a square lattice} 
Here we consider the IMD model with NN repulsion between the same type of particles, 
which is an interesting variant of the original Ziff-Gulari-Barshad's catalytic surface-reaction model \cite{zgb}.
A monomer (denoted by A) adsorbs on a (randomly selected) vacant site with 
free adsorption-attempt probability $P$.
Likewise a dimer (denoted by BB) adsorbs on two (randomly selected) NN vacant sites with 
free adsorption-attempt probability $(1-P)$. 
The parameter $P$ is  the single parameter in the present model.
Adsorbed dimer is assumed to instantly dissociate into two monomers.
In two dimensions, one has to carefully implement the kinetic rules respecting the constraint of NN repulsion:
(a) Monomer A adsorbs on a vacant site only if its NN sites 
are completely empty or have one or more B's regardless of the presence of A. 
Adsorbed A then reacts with one of the NN B's to form the 'molecule' (AB) which leaves the system.
(b) Likewise, dimer BB adsorbs on  two vacant NN sites 
only if their six NN sites satisfy one of the following conditions:  
(1) they are completely vacant.
(2) they have one or more A's in the three left-half (LH) NN sites, 
and at the same time have no B in the three right-half (RH) NN sites, and the vice versa.
(3)  they have  one or more A's on both the LH and RH NN sites, regardless of the presence of B.
The system thus has the two symmetric absorbing states: 
'anti-ferromagnetic (AF)' arrangements of vacant sites and A's.
\begin{figure}[!t]
\begin{center}
\includegraphics[width=1.0\textwidth]{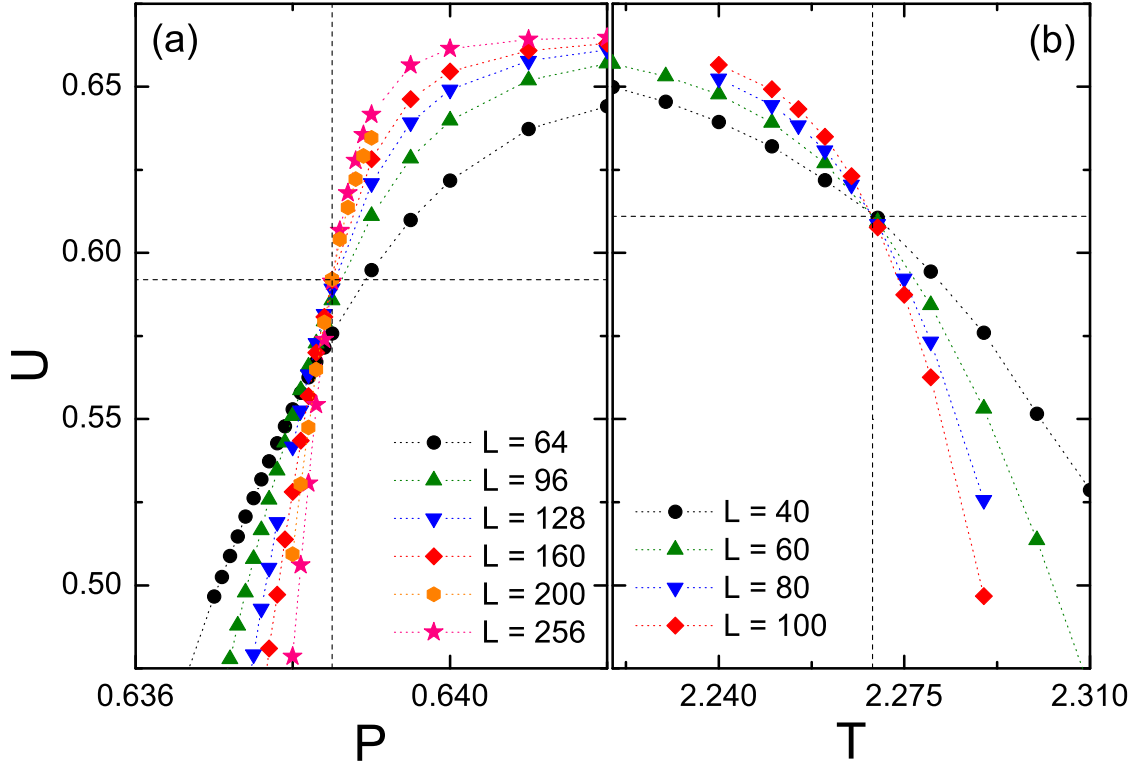}
\caption{The Binder cumulants for the present model((a)) and the EQ Ising model((b)): 
$U(P_{SB})=0.592(1)$ is smaller than the universal value $U_{Ising} \simeq 0.611$ for the 
EQ two dimensional Ising model.}
\end{center} 
\end{figure}

\section{Simulation results and discussions} 
\begin{figure}[t!]
\begin{center}
\includegraphics[width=0.8\textwidth]{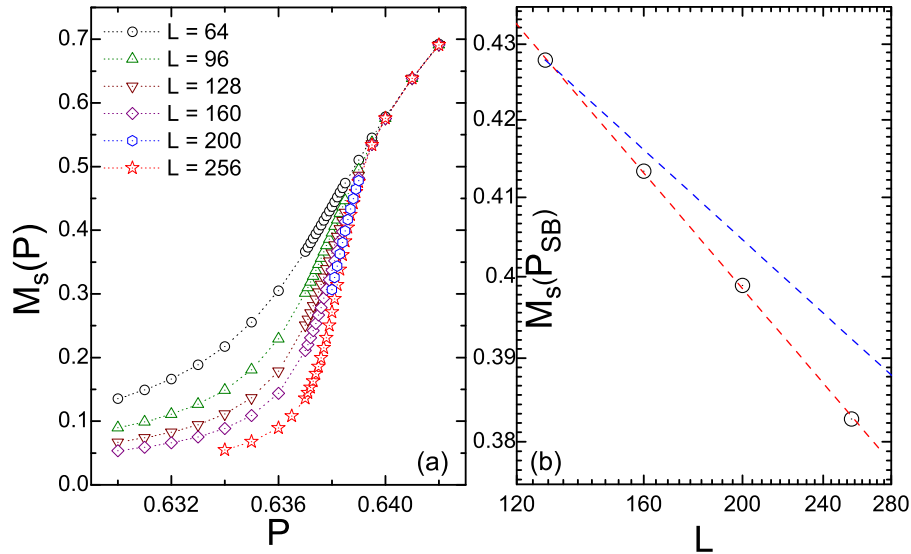}
\caption{(a) The staggered magnetization $M_s(P)$ vs $P$ for various system sizes
 $L$ near the SB transition. 
(b) $M_s$ vs $L$ at the SB transition. 
It shows a power law decrease of  $M_s(P_{SB})$ with $L$ as $M_s(P_{SB}) \sim L^{-\beta/\nu}$ with
$\beta/\nu=0.161(3)$. Ising value $\beta/\nu=0.125$ (blue line) is shown for comparison.
}
\end{center}
\end{figure}

Simulation is performed on a square lattice with linear size $L$ using periodic boundary condition. 
Throughout the work, simulation always starts from the vacuum state.
The 'spin' value is assigned at each site $i$: $\sigma_i= 1, -1, 0$ for A, vacancy, and B, respectively.
Then one can define the local 'staggered magnetization' at site $i$ as 
$m_i = (-1)^{i_x+i_y}\sigma_i$ where $i_{x(y)}$ is the $x(y)$ coordinate of the site $i$. 
The order parameter for the SB transition is given by the staggered magnetization $M_s$, 
 the average of the local staggered magnetization, i.e., 
$M_s = \big< {\cal M}_s \big>$, where ${\cal M}_s \equiv \sum_i m_i/N$ ($N \equiv L^2$).
In order to locate the SB transition point,  
we measure the corresponding Binder cumulant $U$ defined as
$U \equiv 1 -\big< {\cal M}_s^4 \big>/(3 \big< {\cal M}_s^2 \big>^2) $.


\begin{figure}[!t]
\begin{center}
\includegraphics[width=1.0\textwidth]{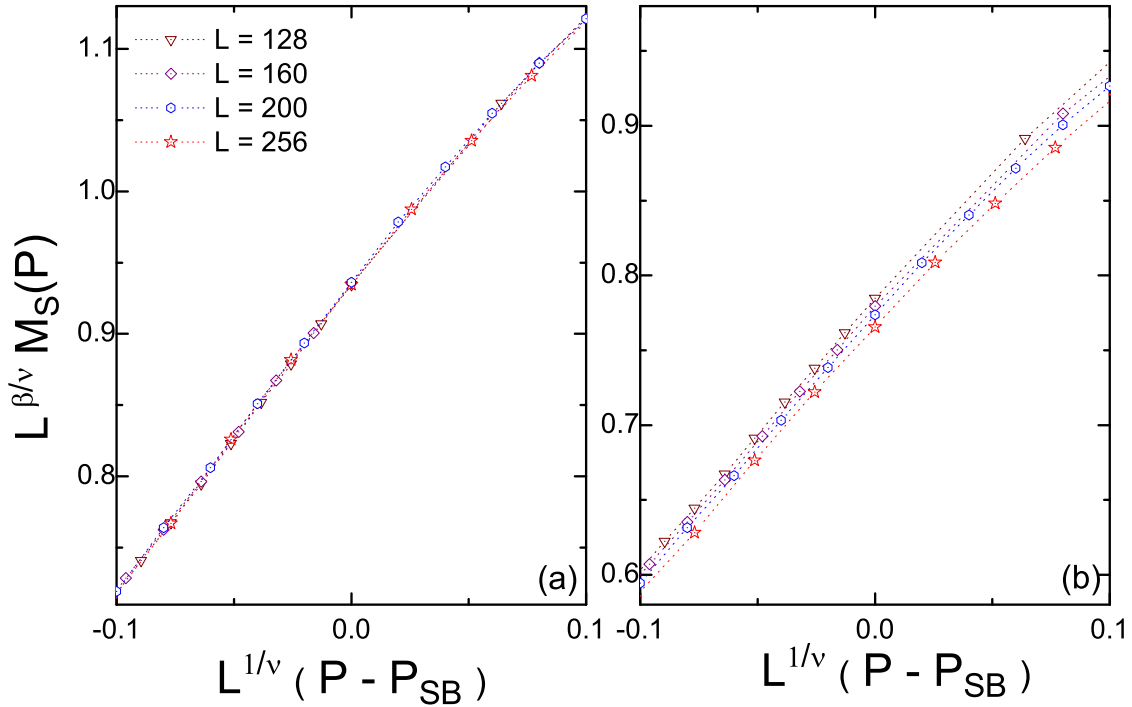}
\caption{Finite size scaling results of $L^{\beta/\nu}M_S$ vs $L^{1/\nu} (P-P_{SB})$ with $\beta=0.161$ and $\nu=1.0$((a)), 
and with the Ising values $\beta=0.125$ and $\nu=1$((b)). }
\end{center}
\end{figure}
Shown in Fig.~1(a) is $U$ as a function of $P$ for various system sizes $L=64 \sim 256$. 
It takes the size-independent universal value $U(P_{SB}) =0.592(1)$ at the transition, 
from which the critical point is determined to be  $P_{SB}=0.6385(1)$.  
For comparison,  we show $U$ for the EQ two-dimensional Ising model  in Fig.~1(b)
which gives  the universal value $U_{Ising} \simeq 0.611$ at $T_c=2.2692$.  
We see that  the observed value $U(P_{SB})$ for the present model
 is found to be {\em smaller} than the pure Ising value. 
We contend that this difference is a manifestation of the non-Ising nature of 
the SB transition. In order to investigate this point further, 
we  measure the critical exponents of the SB transition via 
the finite size scaling (FSS) method and a NEQ kinetic method involving the critical coarsening.

Figure~2(a) shows a plot of $M_s$ vs $P$ for various system sizes.
It is expected that $M_s$ obeys the FSS of the form 
$M_s (P) = L^{-\beta/\nu} F((P-P_{SB}) L^{1/\nu})$. 
Thus its critical value $M_s(P_{SB})$ will vanish wih system size $L$ exhibiting a power law decay  as 
$M_s(P_{SB}) \sim L^{-\beta/\nu}$. Shown in Fig.~2(b) is  
a double-log plot of $M_s(P_{SB})$ vs $L$ whose slope for $L=128,160,200,256$ gives 
 $-\beta/\nu = -0.161(3)$ (Note that the Ising slope $-\beta/\nu = -0.125$ considerably deviates from the actual slope of the data).
With this measured value of $\beta/\nu = 0.161$, we find that the best scaling collapse is obtained for 
$\nu \simeq 1.0$, as shown in Fig.~3(a). 
In contrast, when we use the exact Ising exponents $\beta=0.125$ and $\nu=1$, we see that 
the scaling is systematically violated, as shown in Fig.~3(b). 
Thus the estimated critical exponents are given by $\beta \simeq 0.161$ and $\nu \simeq 1.0$. 

\begin{figure}[!t]
\begin{center}
\includegraphics[width=0.8\textwidth]{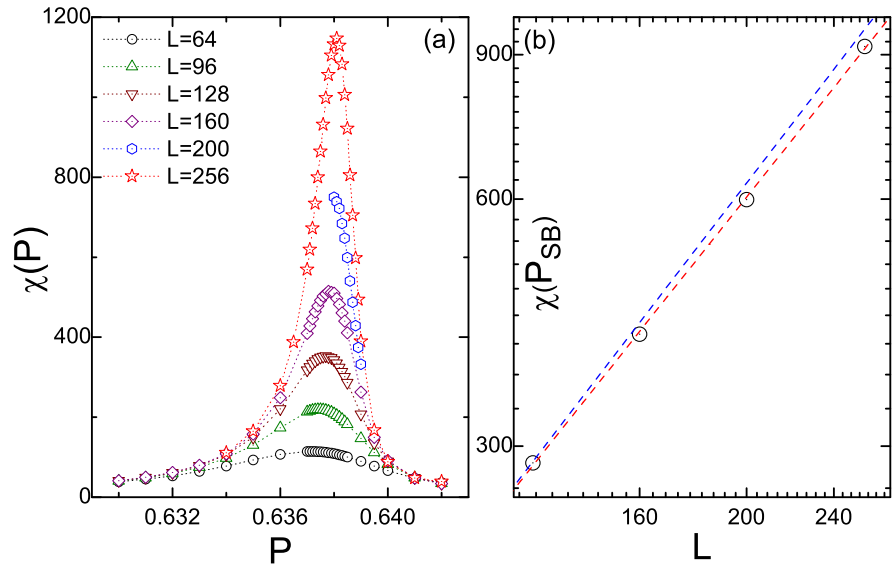}
\caption{(a) Fluctuations of the staggered magnetization 
$\chi$ vs $P$ for different system size $L$.
(b) $\chi$ vs $L$ at the SB transition.
It shows a power law increase with system size $L$ as 
$\chi \sim L^{\gamma/\nu}$ with $\gamma/\nu =1.69(2)$. 
Ising value $\gamma/\nu=1.75$ (blue line) is shown for comparison.}  
\end{center}
\end{figure}

\begin{figure}[t!]
\begin{center}
\includegraphics[width=1.0\textwidth]{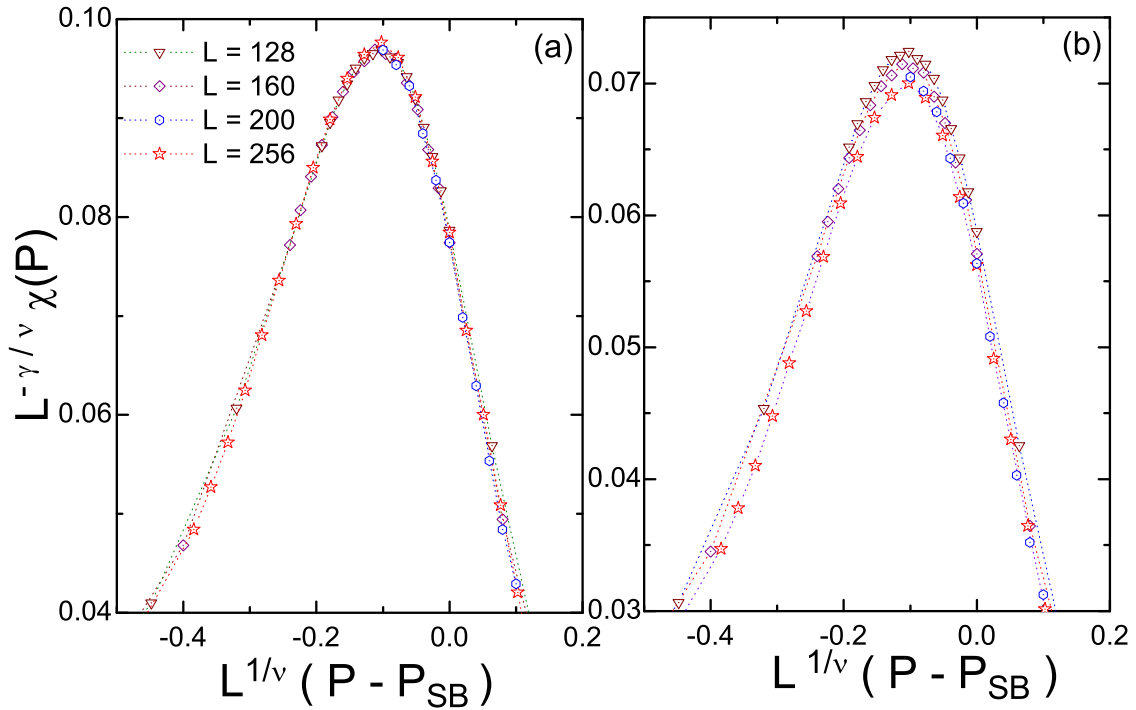}
\caption{
Finite size scaling results of $L^{-\gamma/\nu} \chi$ vs $L^{1/\nu}(P-P_{SB})$ with $\gamma=1.69$ and $\nu =1.0$((a)), 
and with the Ising values $\gamma/\nu=1.75$ and $\nu=1$((b)).
}
\end{center}
\end{figure}

We also measured the fluctuation of the order parameter
$\chi = N \big[ \big< {\cal M}_s^2 \big> - \big< {\cal M}_s \big>^2  \big]$, which
is plotted against $P$ for different system sizes in Fig.~4(a).
It tends to diverge with increasing $L$, and is expected to 
obey the FSS $\chi = L^{\gamma/\nu} G\big(L^{1/\nu}(P-P_{SB}) \big)$. 
Thus, $\chi(P_{SB})$ will diverge with system size as 
$\chi(P_{SB}) \sim L^{\gamma/\nu}$, which is shown in the inset of Fig.~4(b) with
$\gamma/\nu=1.69(2)$ (The Ising slope $\gamma/\nu=1.75$ is shown to deviate from the actual slope of the data).
With this measured value, figure~5 (a) shows that the best scaling is again obtained for $\nu \simeq 1.0$. 
Note again that the scaling breaks down for the Ising value $\gamma=1.75$ and 
$\nu=1$, as demonstrated in Fig.~5(b). 
Therefore, we get $\gamma =1.69(2)$ and $\nu \simeq 1.0$.

To further check the non-Ising nature of the SB transition, we obtain 
the critical exponent $\eta$ from the {\em dynamic} order-parameter correlation function 
$C(r,t)=\big< \sum_i m_i(t) m_{i+r}(t)   \big>/N$ measured during 
the critical coarsening process ensued by a quench from
an initial vacuum state to the SB critical point in a large system with $L=2048$, which does not
involve the data collapse of FSS method. 
 Figure~6 shows $C(r,t)$ for various times.  
At long times, the steady state behavior emerges and grows in the short and intermediate length regions, 
exhibiting a critical decay as $C(r) \sim r^{-\eta}$ with $\eta=0.310(3)$. 
Note that the Ising slope $\eta_{Ising}=0.25$ shown in the figure considerably deviates from the actual data.
We also note that the above estimates of the critical exponents are in good agreement with 
the known scaling relations $2 \beta/\nu=\eta$ (in two dimensions) and $\gamma/\nu=2-\eta$
since $2\beta/\nu=0.322(6)$ and $\gamma/\nu=1.69(2)$.
\begin{figure}[t!]
\begin{center}
\includegraphics[width=0.7\textwidth]{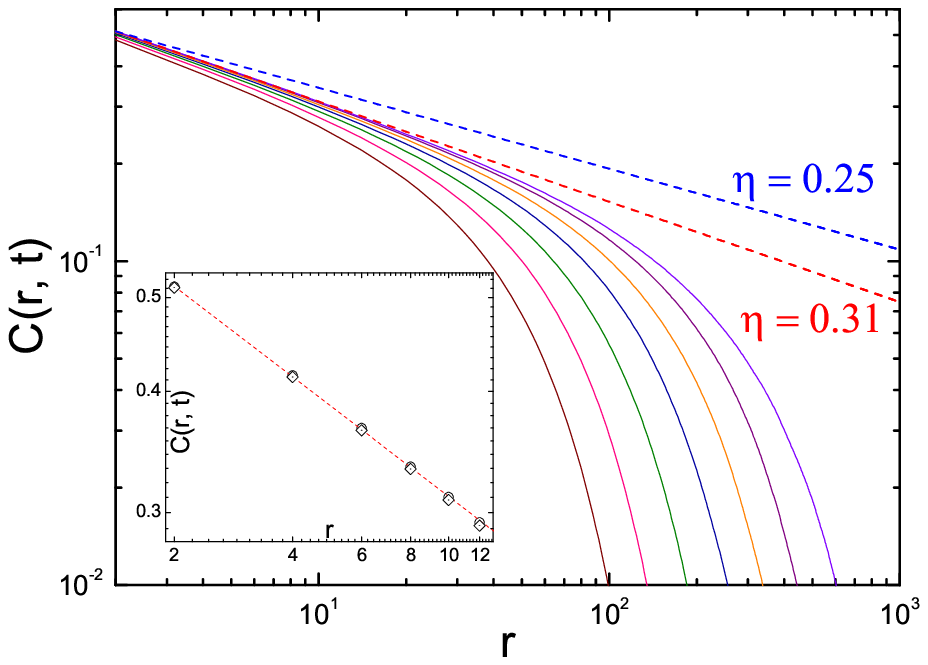}
\caption{The staggered magnetization correlation function $C(r,t)$ vs $r$ for different
times $t=10 \cdot 2^n (n=7,8, \cdots,13)$ measured during the critical coarsening process ($P=P_{SB}$). 
The steady-state part shows a critical decay $r^{-\eta}$ with $\eta=0.31$ (see the inset). 
Power law decay with Ising value $\eta=0.25$ (blue line) is shown for comparison.
Inset: Power law decay of $C(r,t)$ vs $r$ over steady state region for the two latest times, 
$t=40960$ (diamond) and $t=81920$ (circle). 
The exponent $\eta=0.31$ is obtained from the slope measured over the distance up to $r=10$.}
\end{center}
\end{figure}

\begin{figure}[!t]
\begin{center}
\includegraphics[width=0.6\textwidth]{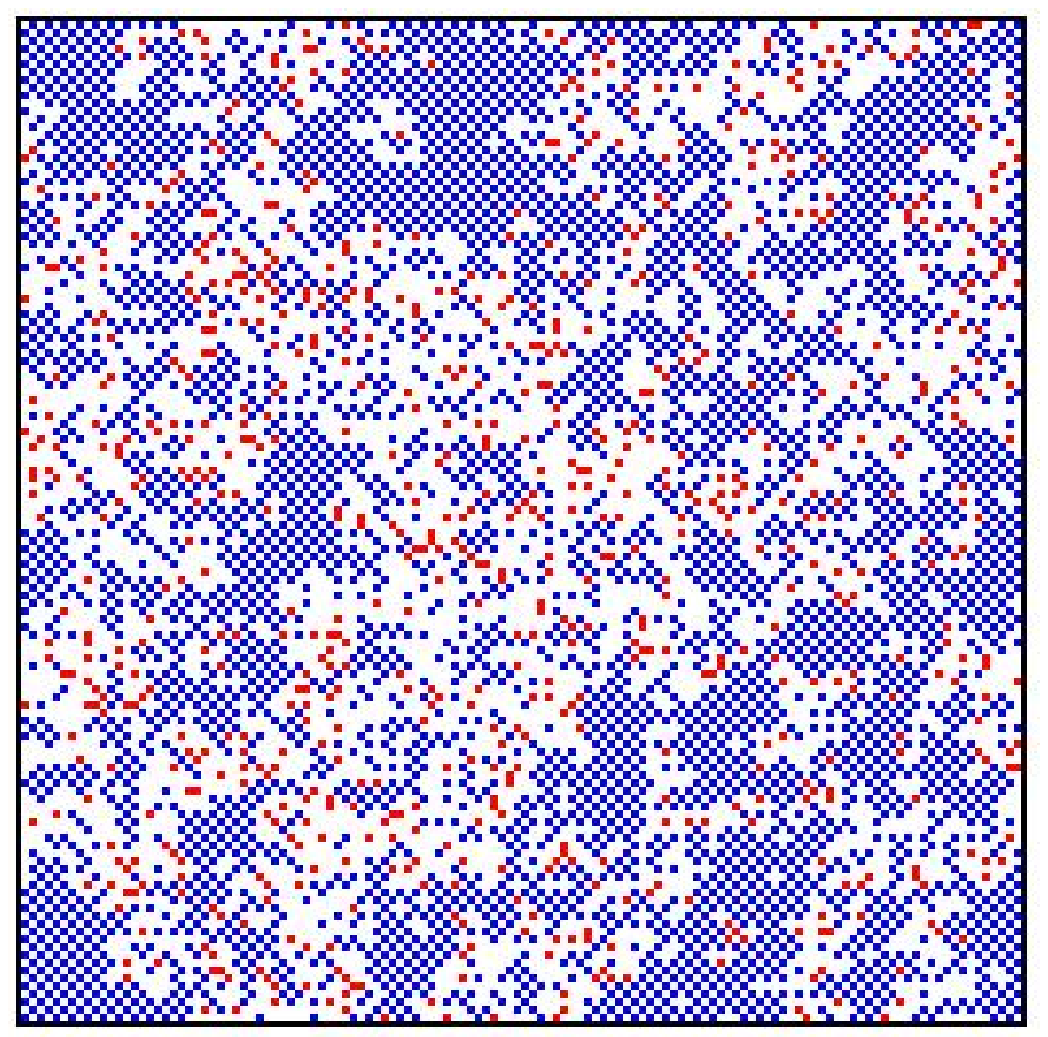}
\caption{A typical steady-state configuration near the SB transition ($P=0.6383$) in a system with $L=128$.
Blue site denotes monomer state (A), red one dimer state (B), and empty one vacant state.}
\end{center}
\end{figure}

Having confirmed the non-Ising criticality of the SB transition,  
we now discuss a possible underlying origin for this non-Ising behavior. 
Although we do not have a solid theoretical argument for the observed non-Ising behavior, 
the AF Ising-like domain morphology appears to provide a crucial qualitative hint for the non-Ising nature of the SB transition 
(see Fig.~7). 
In the present model, the two symmetric absorbing states correspond to the checkerboard pattern of the monomer A's due to
the NN repulsion and NEQ nature of dynamic rules.  One distinct feature in the present model is the presence of 
the dimers. We can see that the existence of the dimers induces (via dynamic rules) longer ranged effective 
repulsion interactions between the monomers, creating wide vacant regions percolating through the system (see Fig.~7).
From the viewpoint of the AF order, these vacant regions contain densely packed defects, and 
must contain many B elements as well within them. These defects as well as B's are expected to divergingly 
fluctuate in their numbers at the SB transition.  Quite probably it is the existence of  these defects and B elements 
(with their diverging fluctuations) that causes the non-Ising critical behavior of the SB transition.
 For example, it is rather clear that these defects and B's can play a crucial role such that  the spatial correlation of the AF order parameter  exhibits more rapid algebraic decay compared to the pure Ising case at the SB transition.
 
 Our preliminary results for the number fluctuation of the defects indeed indicates that it diverges at the SB transition, and its presentation is deferred to a later publication. 
Here we present only the number fluctuations of the B element, i.e., 
$\chi_B = N \big[ \big< \rho_B^2 \big> - \big< \rho_B \big>^2 \big]$ at the SB transition. 
We indeed find that this quantity also exhibits a diverging behavior at the SB transition, 
though much weaker one than the staggered magnetization fluctuations $\chi$, as shown in Fig.~8.
The finite size scaling of $\chi_B$  is given by 
$\chi_B = L^{\gamma_B/\nu_B} H \big( L^{1/\nu_B} (P-P_{SB}) \big)$. 
At the SB transition, the double-log plot $\chi_B$ vs $L$ gives the slope
$\gamma_B/\nu_B = 0.212(3)$. With this value, we find that 
the best scaling collapse is obtained for $1/\nu_B=0.80(5)$, as 
shown in Fig.~8. This gives $\gamma_B =0.265(22) $ and $\nu_B =1.25(8)$.


\begin{figure}[!t]
\begin{center}
\includegraphics[width=0.8\textwidth]{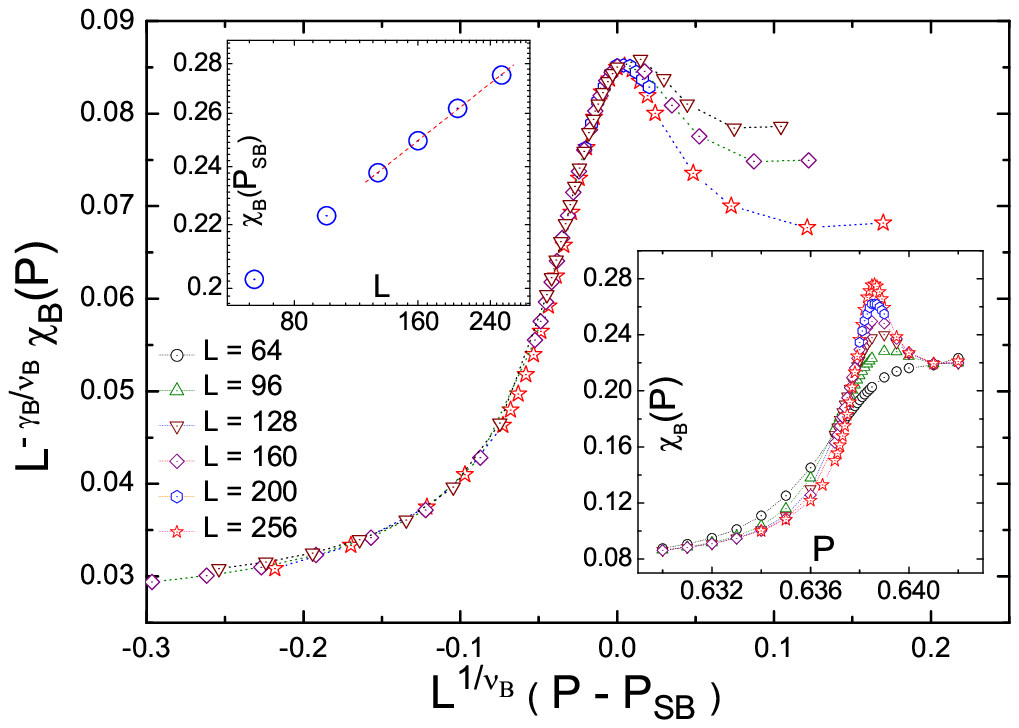}
\caption{Lower inset: The dimer fluctuations 
$\chi_B$ vs $P$ for different system size $L$.
Upper inset:  
$\chi_B$ vs $L$ at the SB transition.
It shows a power law increase with system size $L$ as 
$\chi_B \sim L^{\gamma_B/\nu_B}$ with $\gamma_B/\nu_B =0.212(3)$.  
Main frame: FSS of $L^{-\gamma_B/\nu_B} \chi_B$ vs $L^{1/\nu_B}(P-P_{SB})$
with $\gamma_B=0.265$ and $\nu_B=1.25$.}
\end{center}
\end{figure}

\begin{figure}[t!]
\begin{center}
\includegraphics[width=0.8\textwidth]{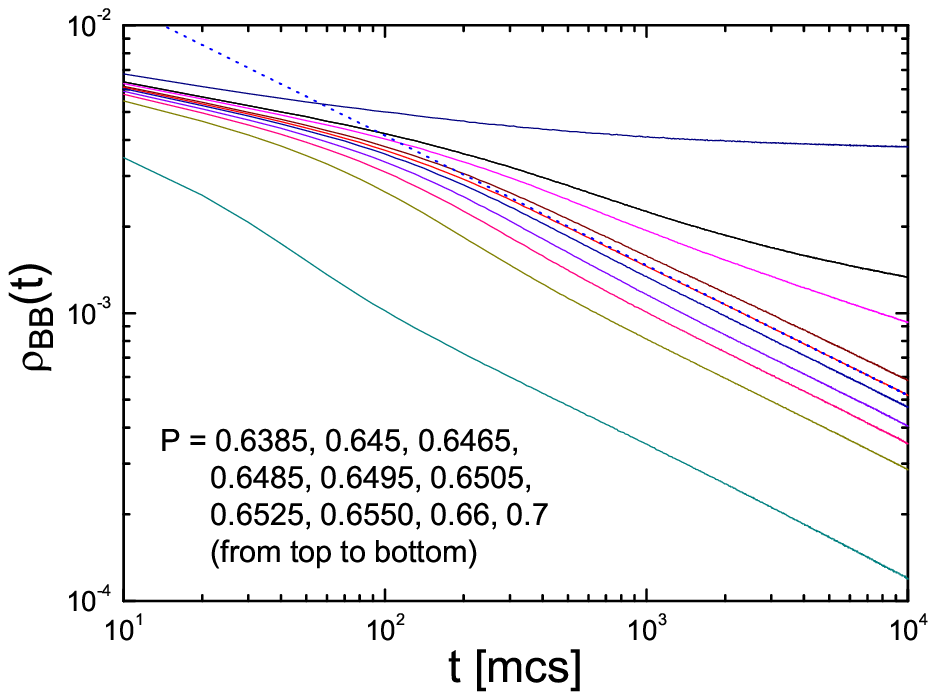}
\caption{The dimer density relaxation $\rho_{BB}(t)$ vs $t$ for various values of $P$ in a system with $L=2048$.
The absorbing transition point is estimated to be $P_A=0.6495(10)$, at which $\rho_{BB}(t)$
exhibits a power law relaxation $\rho_{BB}(t) \sim t^{-\phi}$ with $\phi \simeq 0.45$ (dotted line denotes a line with the slope -0.45). 
The same power-law relaxation is observed to persist within the absorbing phase.} 
\end{center}
\end{figure}
In the present model, the SB and the absorbing transitions
are split, and the former occurs within the active phase (as shown in Fig.~9), 
and the full AF order is achieved only within the absorbing phase. 
By monitoring the relaxation of the dimer density $\rho_{BB}(t)$, the order parameter
of the absorbing transition, for various $P$'s, as shown in Fig.~9, 
we obtain an estimate for the absorbing transition point
as $P_A =0.6495(10)$ where the dimer density $\rho_{BB}(t)$ exhibits 
a power law decay $\rho_{BB}(t) \sim t^{-\phi}$ with $\phi \simeq 0.45$.
This exponent is the same as that in the two-dimensional DP absorbing transition, 
which indicates that the absorbing transition occurring at $P_A$ 
belongs to the DP universality class. 
This behavior is in accord with the expectation \cite{grassberger2,janssen} that 
the the critical behavior is of DP class if the $Z_2$ symmetry of the absorbing state is broken.
Remarkably, the same critical relaxation of the dimer density 
is observed to persist even within the entire absorbing phase ($P > P_A$). 
In this sense, the absorbing phase may be considered as critical.
Note that since $Z_2$ symmetry is already broken for $P \ge P_A $, 
the system evolves toward the absorbing steady-state via coarsening of the AF domains.
The AF domains are absorbing and thus monomers and dimers cannot fall on them.
They therefore must fall on the vacant regions which A's and B's are scattered within.
Coarsening process thus is driven by this  interface-alone-fluctuations, and is expected to 
exhibit a dynamic scaling for the spatio-temporal correlations of the AF order parameter.
We may consider the observed persistent critical decay of the dimer density even within the absorbing phase
as associated with this inevitable coarsening dynamics from initial vacuum state evolving
toward the final (symmetry-broken) absorbing state at {\em all} points of the absorbing
phase. In other words, if we assume that the total number of the dimers is proportional to the volume 
of the vacant regions, then the power law decay of the dimers will ensue from the
power law decay of the volume of vacant regions associated with the kinetic
self-similarity of the system during the coarsening (i.e., phase ordering) process.              
As for the universal DP exponent value in the whole absorbing phase, we may draw some analogy
to the "zero temperature fixed point" in the phase ordering of a standard  Ising model quenched
to a temperature below the transition temperature $T_c$, where the phenomenology of coarsening
at {\em any} temperature below $T_c$ can be essentially represented by the {\em same} 
universality of dynamic criticality.

We now discuss on the continuum stochastic description for the GV universality class.
In Ref.[10], the following Langevin equation for the local SB order parameter field $\phi $ was proposed for a unified description of 
the two types of behaviors alluded earlier for the models in the GV class:
\be
\partial_t \phi =\big( a \phi - b \phi^3 \big) \big( 1 -\phi^2 \big)  +  \nabla^2 \phi + \sigma \sqrt{ 1 -\phi^2} \xi
\la{eq1}
\ee
where $\xi $ is a uncorrelated Gaussian noise with zero mean and unit variance, and $a$,$b$, and $\sigma$ are constants. 
This equation is a modified version  of the Langevin equation for the voter model \cite{dickman,munoz} ((\ref{eq1}) with $a=b=0$)
so as to incorporate the splitting of the $Z_2$ SB transition and the absorbing transition, which was first observed in the model studied in \cite{droz}.
Since the order parameter field fluctuates around $\phi=0$ (and hence $ \big(1-\phi^2 \big)\simeq 1 $) near the SB transition, 
the multiplicative factor in the noise term in (\ref{eq1}) disappears, reducing (\ref{eq1}) to that of the model A.  
This argument seems to be consistent with the report of Ref.[10] that the convincing numerical solution of (\ref{eq1}) reveals that the $Z_2$ SB transition described by (\ref{eq1})
belongs to the Ising universality class, which are evidenced by the FSS scaling and  measurements of the Binder cumulant and time auto correlation function for the order parameter (though these results are not shown explicitly in that paper). 
It is thus quite probable that the Langevin equation (\ref{eq1}) cannot fully describe the non-Ising critical behavior of the SB transition observed in the present model. 

Recently, Vazquez and L\'opez (VL) \cite{vazques} derived a similar Langevin equation from the microscopic model of the Ising spins $\sigma_{\bf r}$
 on a square lattice with spin flip dynamics controlled by a flip probability which is given by a  function $f (\psi) $ of the local magnetization 
$\psi ({\bf r},t) \equiv \big(\sum_Z \  \sigma_{\bf r}\big)/Z$ 
 ($Z$ being the coordination number)  with $\sum_Z$ denoting sum over $Z$ neighboring spins of $\sigma_{\bf r}$. 
This was done by mapping the master equation into an approximate Fokker-Planck equation, and in turn into  the corresponding Langevin equation which is given by
 \bea
 \partial_t \phi &=& \big( a \phi - b \phi^3 \big) \big( 1 -\phi^2 \big)  +  X(\phi) \nabla^2 \phi +  \sqrt{Y(\phi)} \xi,  \nonumber \\
X(\phi)   &\equiv&  \big[ a+c+(d-2a-3b) \phi^2 \big],  \nonumber \\
Y (\phi) &\equiv&  \big(1 -\phi^2 \big) \big( c+d \phi^2 \big)  + \big( a-c+2d \big) \phi \nabla^2 \phi   
\la{eq2}
 \eea
 The corresponding spin-flip probability function is given by $f(\psi) = \big(1 + \psi \big) \big( c+a \psi + d \psi^2 -b \psi^3 \big)/2$ with $a,b,c$ and $d$ being constants. 
 VL found through the simulation on a square lattice that the splitting of the SB and absorbing transitions is observed when the next-next-nearest-neighbor ($Z=12$) spins are taken into 
 account in the calculation of the local magnetization $\psi$, and that the Binder cumulant $U \simeq 0.56$ was obtained at the SB transition.  We tend to believe that this value of $U$, considerably smaller than the pure Ising value $U=0.61$, indicates that the SB transition exhibits a non-Ising critical behavior. 
 It is interesting to see that unlike (\ref{eq1}), the corresponding Langevin equation (\ref{eq2}) does not seem to reduce to the model A near the SB transition due to the forms of $X(\phi)$ and $Y(\phi)$ in (\ref{eq2}).
 This feature could leave a room for the possibility that the Langevin equation (\ref{eq2}) might be able to generate a non-lsing critical behavior of the SB transition, which deserves a thorough investigation on the critical behavior of the SB transitions in both microscopic model and  corresponding Langevin equation proposed by VL. 
 
 \section{Summary}
 In the present work, we investigated a two dimensional IMD model with short-range repulsion, which has
 two symmetric absorbing states. 
We find that the SB and absorbing transitions take place at two nearby points, 
as in some models \cite{droz,vazques} with extended interaction range belonging to the GV universality class. 
Interestingly, we observe numerically that the SB transition exhibits a new non-Ising critical behavior 
with the exponents $\beta \simeq 0.161$, $\gamma \simeq 1.69$, $\nu \simeq 1.0$, and $\eta \simeq 0.31$.
We also find that the absorbing transition is of the DP-type, with a remarkable feature that the dimer density shows the same critical 
decay (with the DP-exponent) within the entire absorbing phase. 
The non-Ising critical behavior of the SB transition is considered to be closely related to the fact that 
the effective extended range of repulsion between  monomers  mediated by dimers (via NEQ dynamic rules) 
creates wide vacant regions between the AF domains, which can be considered as closely packed defects from the view of the AF order. We suggested that the diverging fluctuations of these defects and dimers can lead to a non-Ising critical behavior of the AF order parameter at the SB transition.  The persistent criticality in the absorbing phase can be attributed to 
the self-similar nature of the coarsening  process (i.e., the AF domain growth) evolving toward symmetry-broken absorbing state.
Finally, we raise a possibility that a recently derived Langevin equation based on the spin-flip dynamics of a collection of Ising spins on
a square lattice with a general flip probability 
might be able to describe similar non-Ising critical behavior of the SB transition. 
It would thus be valuable to carry out further analysis on those models and continuum stochastic equations showing 
separate transitions in the GV universality class.
It would be also interesting to conduct more detailed investigations into NEQ kinetics 
in both active and absorbing phases in the present model. 

We thank Hyunggyu Park for valuable discussions.  
This work was supported by a National Research Foundation of Korea (NRF) grant funded 
by the Korean government (MEST: No. 2009-0090085) (SP and SJL), and by Changwon National University Grants 2009 and 2010 (BK).
The computation of this work was supported by PLSI supercomputing resources of Korea Institute of  Science and Technology Information (KISTI). 



\begin{thebibliography}{99}
\bibitem{marro-dickman}
J. Marro and R. Dickman, {\em Non-Equilibrium Phase Transitions in Lattice Models}
(Cambridge University Press, Cambridge, England, 1996).

\bibitem{henkel}
  M. Henkel, H. Hinrichsen, S. L\"ubeck, 
{\em Non-Equilibrium Phase Transitions: Volume 1: Absorbing Phase Transitions}
(Springer, New York, 2009).

\bibitem{odor}
G. \'{O}dor, {\em Universality in Non-Equilibrium Lattice systems}
 (World Scientific, Singapore, 2008).

\bibitem{grassberger1}
P. Grassberger and A. de la Torre,  Ann. Phys. (NY) {\bf 122}, 373 (1979).

\bibitem{hinrichsen} H. Hinrichsen,  Adv. Phys. {\bf 49}, 815 (2000); Physics {\bf 2}, 96 (2009).

\bibitem{takeuchi} K. A. Takeuchi, M. Kuroda, H. Chat\'e, and M. Sano, \PRE {\bf 80}, 051116 (2009).


\bibitem{janssen}
H. K. Janssen, Z. Phys. B {\bf 42}, 151 (1981).

\bibitem{grassberger2}
P. Grassberger, Z. Phys. B {\bf 47}, 365  (1982).



\bibitem{dornic} I. Dornic, H. Chat\'{e}, J. Chave, and H. Hinrichsen, \PRL {\bf 87}, 045701 (2001).

\bibitem{hammal} O. Al Hammal, H. Chat\'{e}, I. Dornic, and M. A. Mu\~{n}oz, \PRL {\bf 94}, 230601 (2005).

 
 \bibitem{castellano} C. Castellano, S. Fortunato, and V. Loreto,  \RMP {\bf 81}, 591 (2009).
 
 \bibitem{menyhard}  N. Menyh\'{a}rd and G. \'{O}dor,  \BJP {\bf 30}, 113 (2000).

\bibitem{nam}
K. Nam, B. Kim, S. J. Lee, J. Stat. Mech. P08013 (2010).

 
 

%

\bibitem{hpark} M. H. Kim and H. Park, \PRL {\bf 73}, 2579 (1994); 
W. Hwang, S. Kwon, H. Park, and H. Park,  \PRE {\bf 57}, 6438 (1998).
%




%

  
 \bibitem{mylee} M. Y. Lee and T. Vojta, \PRE {\bf 83}, 011114 (2011).
 
\bibitem{krapivsky} L. Frachebourg and P. L. Krapivsky, \PRE {\bf 53}, R3009 (1996).


\bibitem{droz} M. Droz, A. L. Ferreira, and A. Lipowski,  \PRE {\bf 67}, 056108 (2003).


\bibitem{vazques} F. Vazques and C. L\'{o}pez,  2008, \PRE {\bf 78}, 061127 (2008).

\bibitem{castellano2} C. Castellano, M. A. Mu\~{n}oz, R. Pastor-Satorras,  \PRE {\bf 80}, 041129 (2009).

\bibitem{blote} G. Kamieniarz and H. W. Bl\"ote, J. Phys. A {\bf 26}, 201 (1993).

\bibitem{zgb} 
R. M. Ziff, E. Gulari, and Y. Barshad, Phys. Rev. Lett. 56, 2553 (1986).

\bibitem{dickman} 
R. Dickman and A. Yu. Tretyakov, \PRE {\bf 52}, 3218 (1995).
\bibitem{munoz}
M. A. Munoz, G. Grinstein, and Y. Tu, \PRE {\bf 56}, 5101 (1997).






























\end{thebibliography}
\end{document}